\documentclass[12pt]{iopart}
\usepackage[detect-all]{siunitx}
\usepackage[pdftex]{graphicx}
\usepackage{xspace}
\usepackage{tabularx}
\usepackage{cite}
\usepackage{color}
\newcommand{\bite}{Bi$_{2}$Te$_{3}$}

\newcommand{\bise}{Bi$_{2}$Se$_{3}$\xspace}
\newcommand{\nb}{Nb\xspace}
\newcommand{\he}{He\xspace}
\newcommand{\icrn}{$I_{\textrm{c}} R_{\textrm{N}}$\xspace}

\begin{document}

\title[Geometric dependence of Nb-Bi${_2}$Te${_3}$-Nb topological Josephson junctions ...]{Geometric dependence of Nb-Bi${_2}$Te${_3}$-Nb topological Josephson junction transport parameters}

\author{CG Molenaar$^1$, DP Leusink$^1$, XL Wang$^2$ and A Brinkman$^1$}

\address{$^1$ Faculty of Science and Technology and MESA+ Institute for Nanotechnology, University of Twente, 7500 AE Enschede, The Netherlands}
\address{$^2$ Institute for Superconducting and Electronic Materials,
University of Wollongong, Wollongong, NSW, 2522, Australia}
\ead{c.g.molenaar@utwente.nl}
\begin{abstract}
Superconductor-topological insulator-superconductor Josephson junctions have been fabricated in order to study the width dependence of the critical current, normal state resistance and flux periodicity of the critical current modulation in an external field. Previous literature reports suggest anomalous scaling in topological junctions due to the presence of Majorana bound states. However, for most realised devices, one would expect that trivial $2\pi$-periodic Andreev levels dominate transport. We also observe anomalous scaling behaviour of junction parameters, but the scaling can be well explained by mere geometric effects, such as the parallel bulk conductivity shunt and flux focusing. 
\end{abstract}

\maketitle

\section{Introduction}

Topological insulator (TI) – superconductor (S) hybrids are potential systems for realizing p-wave superconductivity and hosting Majorana zero-energy states \cite{Read2000,Kitaev2001,Ivanov2001,Fu2008,Nilsson2008,Tanaka2009,Stanescu2010,Potter2011}. The common singlet s-wave pairing from a nearby superconductor is predicted to induce a spinless p-wave superconducting order parameter component in a topological insulator \cite{ Fu2008, Potter2011} because of the spin-momentum locking of the surface states in a topological insulator \cite{Hsieh2009a,Chen2009,Hasan2010}. In a Josephson junction between two s-waves superconductors with a topological insulator barrier (S-TI-S), Majorana Bound States (MBS) can occur with a $\sin(\phi/2)$ current-phase relationship \cite{Fu2008}. Contacts between superconductors and 3D topological insulators were realised on exfoliated flakes and films of \bite{}, \bise{}, and strained HgTe \cite{Sacepe2011,Veldhorst2012,Yang2012,Williams2012,Oostinga2013,Cho2013,Galletti2014}, and the Josephson behaviour was investigated by measuring Fraunhofer patterns in the presence of an applied magnetic field and Shapiro steps due to microwave radiation \cite{Veldhorst2012,Yang2012,Williams2012,Cho2013,Galletti2014}. Despite the presence of conductivity shunts through bulk TI, the Josephson current was found to be mainly carried by the topological surface states \cite{Veldhorst2012,Cho2013,Galletti2014}.

Pecularities in the Fraunhofer diffraction patterns have been found for topological Josephson junctions \cite{Williams2012, Kurter2014}, including non-zero minima in the Fraunhofer patterns and periodicities which do not correspond to the junction size. In junctions with a varying width the characteristic energy \icrn was reported to scale inversely with the junction width \cite{Williams2012}. This observation has been phenomenologically attributed to the width dependence of the Majorana modes contributing to a highly distorted current-phase relationship \cite{Williams2012}.  The Majorana modes have also been held responsible for the unexpectedly small flux periodicities in the $I_c(B)$ Fraunhofer pattern of the same junctions \cite{Williams2012}. 

However, only one mode out of many channels is a MBS. For all non-perpendicular trajectories a gap appears in the Andreev Bound state spectra, giving trivial $2\pi$ periodic bound states \cite{Snelder2013}. For typical device sizes fabricated so far, the number of channels is estimated to be large (a width of the order of a few 100 nm and a Fermi wavelength of the order of $k_F^{-1}=1$ nm already gives a few 100 modes). The Majorana signatures are, therefore, expected to be vanishingly small.

To understand the $I_c(B)$ periodicity as well as the scaling of \icrn with width, we have realized S-TI-S topological Josephson junctions with varying width. We also observe a non-trivial scaling of the critical current, normal state resistance and magnetic field modulation periodicity. However, a detailed analysis shows that all scaling effects can be explained by mere geometric effects of trivial modes. The dominance of trivial Andreev modes is supported by the absence of $4\pi$ periodicity signatures in the Shapiro steps under microwave irradiation.

\section{Expected Majorana related modifications of the critical current modulation by magnetic field and microwaves}
Screening external flux from a superconducting junction results in the characteristic Fraunhofer pattern in Josephson junctions due to the DC Josephson effect. The critical current is modulated by the magnetic flux with a periodicity of the superconducting flux quantum, $\Phi_0=h/2e $, threading the junction, due to the order-parameter being continuous around a closed contour. If the current-phase relationship is changed from $\sin(\phi)$ to $\sin(\phi/2)$ in a topologically non-trivial junction the periodicity is expected to become $h/e$. 

Additionally, for junctions where MBS are present it has been proposed that the minima in the Fraunhofer pattern are non-zero \cite{Potter2013}. The current at the minima is predicted to be approximately equal to the supercurrent capacity of a single channel, $I_M \approx \Delta/\Phi_0$.

Applying an AC bias on top of the DC bias will create a frequency to voltage conversion, the AC Josephson effect. In the voltage state of the junction, at DC voltages equal to $k \Phi_0 f = k h f / 2e $, with $k$ an integer and $f$ the frequency in Hz, there will be a current plateau with zero differential resistivity at fixed finite voltages. The presence of these Shapiro steps in a superconducting junction is one of the hallmarks of the Josephson effect. In contrast to a Fraunhofer pattern, it does not depend on the geometry of the junction, but on the current-phase relationship of the junction. A sum of different current-phase relationships \cite{Kwon2003,Kwon2004,Kwon2004a,Fu2009,Ioselevich2011,Pal2014} $I = A_1 \sin{\phi/2} + A_2 \sin{\phi} + A_3 \sin{2\phi} + \ldots$ will result in current plateaus at $V_{1,l} = l h f/e$, $V_{2,m} = m hf/2e$, $V_{3,n} = n hf/4e$, etc. For a pure $\sin(\phi/2)$ relationship one expects steps only at $ k hf / e $. The actual width and the modulation as function of applied RF power of the current plateaus depends on the ratio between the applied RF frequency and the \icrn{} product of the junctions. This can be numerically obtained by solving the Resistively Shunted Josephson (RSJ) \cite{Tinkham2004} junction model.

\section{Sample layout and fabrication}
Devices were designed with junctions of constant electrode separation and varying width. The fabrication is similar to the method used by Veldhorst \textit{et al.}\ \cite{Veldhorst2012}, but has been modified to reduce the number of fabrication steps and increase the number of usable devices available on one chip. Exfoliated flakes are transferred to a Si/SiO$_2$ substrate. E-beam lithography, with \SI{300}{nm} thick PMMA resist, is used to define junctions and contacts in two different write fields, eliminating the photo-lithography step.

\begin{figure} 
	\centering
		\includegraphics[width=1\linewidth]{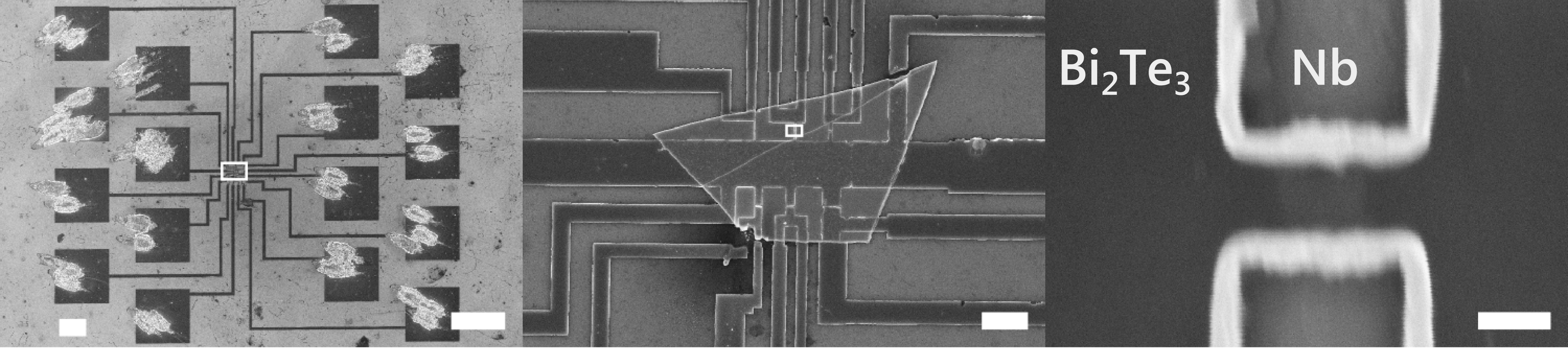}
		\caption{ \textsc{Scanning electron microscope images of a typical device.} The white bar is \SI{200}{\micro m}, \SI{5}{\micro m} and \SI{100}{nm} wide in the three consecutive images respectively, and the white rectangles in the two left images mark the location of the image to the right. The first image show the Nb contact pads (dark grey) written with a large write field. The middle image shows the flake (trapezoid with bright edges and a step edge diagonal across) with leads (dark grey) leading to the junctions. The junctions are visible as faint white breaks in the leads. Along the top-row, left to right, are a 100, 250 and \SI{500}{nm} junction. On the bottom-row are a 750, 1000 and \SI{2000}{nm} junction. The \SI{5}{\micro m} on the right hand side of the flake is overexposed. The rightmost image is a close-up of a \SI{250}{nm} by \SI{150}{nm} designed junction.}
		\label{fig:bi2te3junctiondesign}
		\end{figure}

In figure~\ref{fig:bi2te3junctiondesign} the contact pads, written with a coarse write field, and the structure on the \bite{} flake, written with a smaller and accurate write field, are visible. The smaller write field increases the resolution possible. An overlap of the structures was used in areas where the dose or write field was changed. These overlaps will cause overexposure and are only possible when the resolution is not critical. The \SI{80}{nm} niobium superconducting film  and a \SI{2.5}{nm} capping layer of palladium are sputter deposited. The flake is Ar-ion etched at \SI{50}{eV} for \SI{1}{min} prior to deposition resulting in transparent contacts.

The edge of the flake with the substrate provides a step edge for the \nb{} from the substrate to the flake, and it is advisable to keep the thickness of the flake comparable or less than the thickness of the \nb{} layer. The thickness of the sputter deposited \nb{} is limited by the thickness of the e-beam resist layer. For flakes of usable lateral dimensions, the thickness is generally in the order of \SI{100}{nm}. The contacts for the voltage and current leads are split on the flake: if a weak link occurs on the lead transition from flake to substrate this will not influence the measured current-voltage ($IV$) characteristic of the junction.

Structures with a disadvantageous aspect ratio (junction width to electrode separation), such as wider junctions, are prone to overexposure. This increases the risk of the junction ends not being separated. For wider junctions a slightly larger separation has been used. Overexposure will decrease the actual separation. Actual dimensions are verified after fabrication as in figure~\ref{fig:bi2te3junctiondesign}.

The junctions are characterised in a pumped \he cryostat with mu-metal screening and a superconducting \nb can surrounding the sample. The current and voltage leads are filtered with a two stage RC filter. A loop antenna for exposure to microwave radiation is pressed to the backside of the printed circuit board (PCB) holding the device. A coil perpendicular to the device surface is used to apply a perpendicular magnetic field. For different values of the applied microwave power or magnetic field current-voltage traces are recorded.

\section{Measured scaling of transport parameters}
\subsection{Junction overview}
The devices are characterised by measuring their $IV$ curves at \SI{1.6}{\kelvin} under different magnetic fields and microwave powers. The microwave frequency of \SI{6}{\giga \hertz} is chosen for maximum coupling as determined by the maximum suppression of the supercurrent at the lowest power. The main measured junction parameters are given in table \ref{tab:bi2te3junctioncharacteristics}.

\begin{table}
	\begin{indented}
	\item[]\begin{tabular}{@{}lll}
	\br
	Junction width (\si{\nano \meter}) & Critical current (\si{\micro \ampere}) & Normal state resistance (\si{\ohm})\\
	\mr
	100  & 0.2 & 1.5 \\ 
	250  & 3.5 & 1.14 \\ 
	500  & 13.5 & 0.84 \\
	1000 & 16 & 0.64 \\ 
	\br
	\end{tabular}
	\end{indented}
	\caption{\label{tab:bi2te3junctioncharacteristics}\textsc{Junction characteristics.} The critical current $I_C$ and normal state resistance $R_N$ are given at \SI{1.6}{\kelvin}. The measured junction separation is  $~$\SI{140}{nm}. The \SI{750}{\nano \meter} and \SI{5000}{\nano \meter} wide junctions have shorted junctions due to e-beam overexposure, and the \SI{2000}{nm} wide junction had a non-ohmic contact, caused by a break at the edge of the \bite{} flake.}
\end{table}

\begin{figure}
	\centering 
	\includegraphics[width=\textwidth]{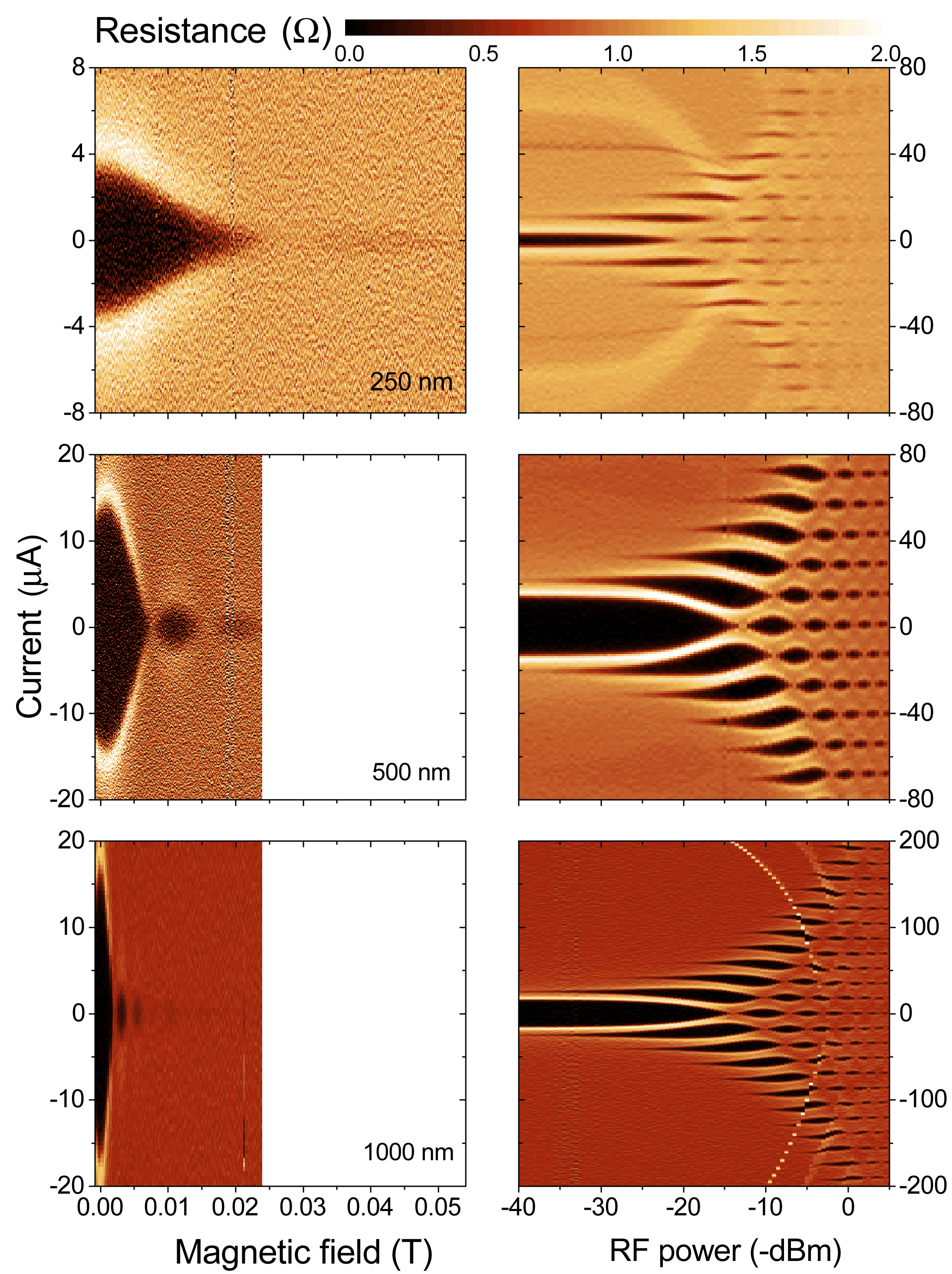}
		\caption{ \textsc{Magnetic field and microwave power dependence.} The top row figures show the dynamic resistance of the \SI{250}{nm} wide junction, the middle row figures correspond to the \SI{500}{nm} wide junction and the bottom row figure to the \SI{1000}{nm} wide junction. The left column shows the reaction to an applied magnetic field, the right column the reaction to microwave power at \SI{6}{GHz}. The horizontal line at \SI{\sim 19}{mT} is an artifact of the magnet current source. $IV$ curves for these junctions are shown in figure \ref{fig:bi2te3Flux}.} 
		\label{fig:bi2te3MagneticAndMicrowaveField}
\end{figure}

Both the magnetic and microwave field response has been studied for all junctions.
Results for the \num{250}, \num{500} and \SI{1000}{nm} wide junctions are shown in figure \ref{fig:bi2te3MagneticAndMicrowaveField}.
The supercurrent for the \SI{100}{nm} wide junction was suppressed in a magnetic and microwave field without further modulation.
In the response to the microwave field a sharp feature is visible starting at \SI{200}{\micro A} and \SI{-10}{dBm} for the \SI{1000}{nm} junction.
This is likely the result of an unidentified weak link in one of the leads.
The fainter structures in the \SI{250}{nm} junction starting at \num{ 43} and \SI{60}{\micro A} and \SI{-40}{dBm} are reminiscent of an echo structure  described by Yang \textit{et al.}\cite{Yang2012} for Pb-\bise{}-Pb Josephson junctions.
Measuring the microwave response at the minima of the Fraunhofer pattern \cite{Potter2013} yielded no Shapiro features.

\subsection{Scaling of $I_\textrm{C}$ and $R_\textrm{N}$}
In general, the normal state resistance, $R_\textrm{N}$, of a lateral SNS junction \cite{Likharev1979,Golubov2004}  is expected to scale inversely with junction width, whereas $I_c$ is expected to be proportional to the width, such that the \icrn product  is constant \cite{Tinkham2004}. Josephson junctions on topological insulators are similar to SNS junctions with an induced proximity effect by superconducting leads into a TI  surface state. For junctions on \bite{} the transport was found to be in the clean limit, with a finite barrier at the interface between the superconductor and the surface states \cite{Veldhorst2012}. The supercurrent for ballistic SNS junctions with arbitrary length and barrier transparency is given by \cite{Galaktionov2002}, which was found to fit the data of Veldhorst \textit{et al.} well \cite{Veldhorst2012}. The normal state resistance in \bite{} is complicated by the diffusive bulk providing an intrinsic shunt. The leads on the \bite{} flake leading up to the junction also contribute towards a normal state conductivity shunt without carrying supercurrent. This results in current paths not only directly between the two electrodes but also through and across the whole area of the flake to the left and right of the electrodes.

The scaling of $I_c$ and $R_N$ with junction width is shown in Figure 3. In the junctions with an aspect ratio (width of the junction divided by electrode separation) greater than 5, the \icrn product is approximately \SI{11}{\micro V}, similar to junctions where the the length was varied instead of the width \cite{Veldhorst2012,Veldhorst2012a}.  Below this, the \icrn product falls sharply. To verify whether this can be due to Majorana modes we estimate the number of conducting channels in these small junctions. The number of channels in a junction is related to the width of the junction: $M = \frac{k_F \times W}{\pi}$~\footnote{The wave vector for linear dispersion is given by $k_F = \frac{E_F}{\hbar v_F} \approx$ \SI{2e9}{m^{-1}}. The Fermi energy is taken as \SI{150}{\milli eV} and the Fermi velocity is in the order of \SI{1e5}{m/s}.}. For a \SI{100}{\nano m} junction this means that there are more than 60 channels active in the junction, and a MBS will not dominate transport properties.

Rather, due to the open edges of the junctions, the scaling of the normal state resistance is not directly proportional to the junction width, but offset due to the whole flake providing a current shunt. This is similar to an infinite resistor network \cite{Cserti2000} providing a parallel resistance to the resistance due to the separation of the two leads. Taking this into account, the resistance between the two leads takes the form $R = (\rho_{\textrm{W}} R_{\textrm{parallel}}) / ( WR_{\textrm{parallel}} + \rho_{\textrm{W}})$, where $rho_{\textrm{W}}/W$ gives the junction resistance without a current shunt and $R_{\textrm{parallel}}$ is the resistance due to the current shunt through the flake. This equation does not disentangle the surface and bulk contributions but treats them as scaling the same. In the zero-width limit, the resistance is cut-off and does not diverge to infinity. The resulting \icrn product can then be well explained by the scaling of $R_N$ (including the shunt) and the usual scaling of $I_c$ with width ($I_c$ being directly proportional to the number of channels, given by the width of the junctions with respect to the Fermi wavelength). Note, that the expected scaling of $I_c$ contrasts previous observations of inverse scaling \cite{Williams2012}.

\begin{figure}
	\centering 
		\includegraphics[width=\textwidth]{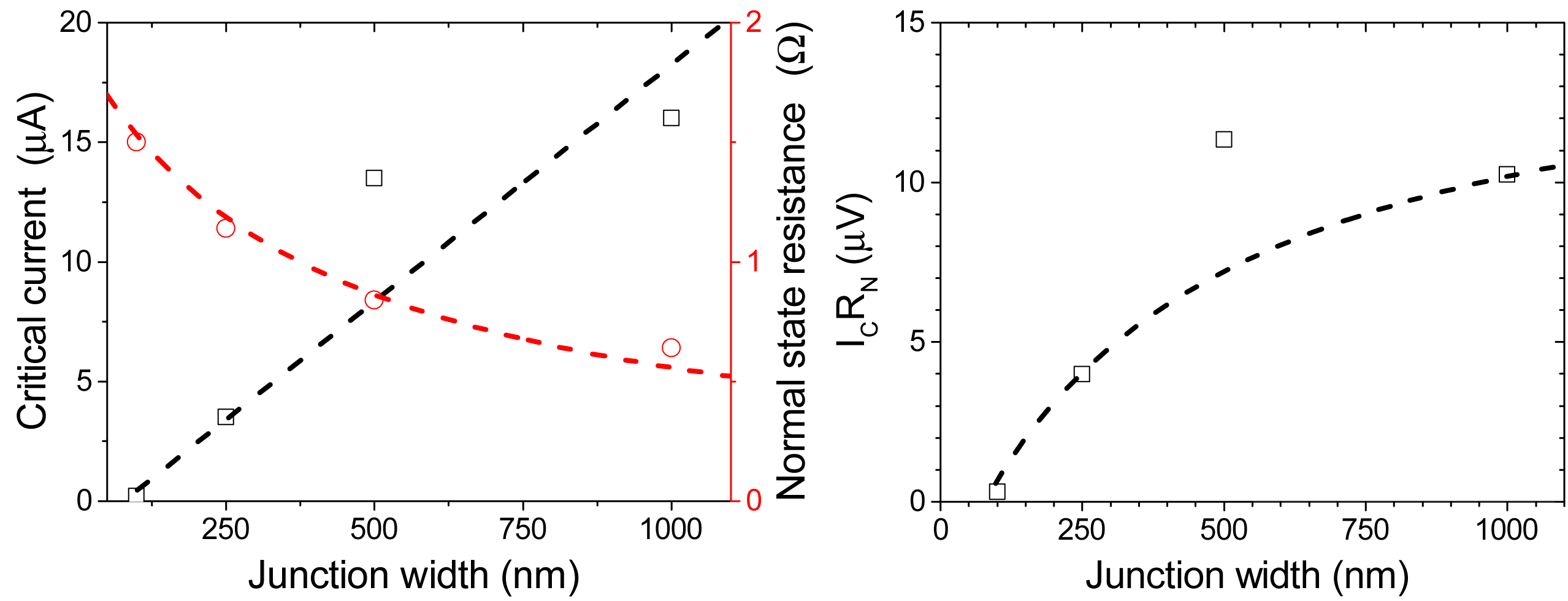}
  		\caption{ \textsc{Scaling of the critical current, normal resistance and \icrn product.} In the left panel the measured critical currents (black squares) and normal state resistances (red circles) are plotted. The critical current is assumed to be linear (black dashes). The resistance is modelled as a width resistivity $\rho_w=$ \SI{7.9e-7}{\ohm m} in parallel  with a constant shunt resistance $R_{\textrm{parallel}}=$ \SI{1.9}{\ohm}. The \icrn product is plotted in the right panel, with the dashed line as the product of the fits in the  left panel. The fit approaches the \icrn product of \num{10} to \SI{15}{\micro V} found in junctions of varying width \cite{Veldhorst2012}.}
  		\label{fig:bi2te3IcandRnandIcRn}
\end{figure}

\subsection{Periodicty in $\Phi_0$}

The critical current of Josephson junctions oscillate in an applied magnetic field due to a phase difference induced across the junction. The magnetic flux in the junction area is the product of the area of weak superconductivity between the two electrodes and flux density in this area. The area of the junction is given by $W \times \left( l + 2\lambda_L \right)$, where $W$, $l$ and $\lambda_L$ are the width, length and London penetration depth respectively. The investigated junctions are smaller than or comparable to the Josephson penetration depth, $\lambda_J = \sqrt{ \Phi_0/( 2 \pi \mu_0 d' J_C) }$, where $d'$ is the largest dimension (corrected by the London penetration depth) of the junction and $J_C$ has been estimated using the bulk mean free path of \bite crystals, \SI{22}{nm} \cite{Veldhorst2012}, which allows us to ignore the field produced by the Josephson current. For the \SI{80}{nm} thick Nb film used we use the bulk London penetration depth, \SI{39}{nm} \cite{Gubin2005, Maxfield1965}.

\begin{figure}
	\centering 
		\includegraphics[width=\textwidth]{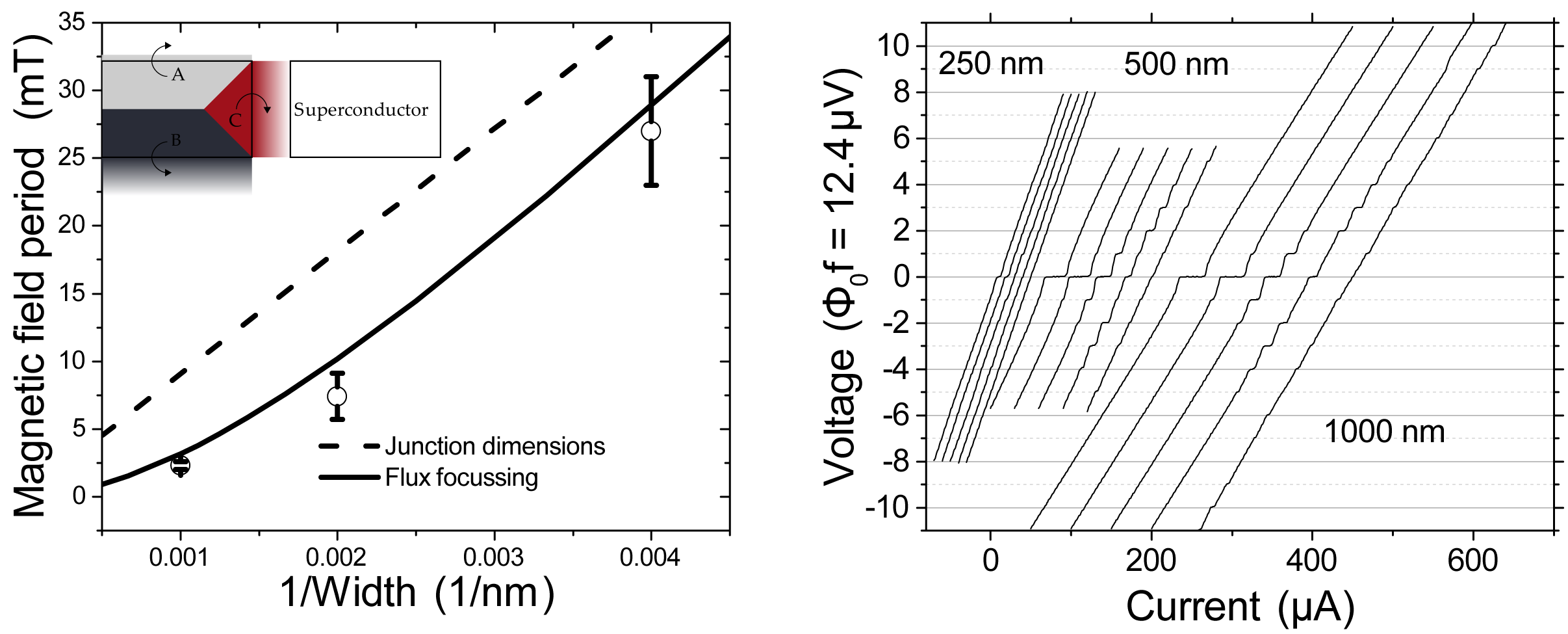}
		\caption{ \textsc{Behaviour of Fraunhofer oscillation frequency and Shapiro steps.} In the left panel the modulation period of the Josephson current as a function of the external field is plotted as a function of the inverse junction width. The dashed line is the expected period for a rectangular junction. The solid line takes into account flux focussing, as presented in the inset. The flux incident on the grey areas A and B is diverted to the sides of the junction lead, while the red area C is added to the effective junction area between the two leads. The dimensions of the superconductor have been reduced by the London penetration depth, since flux can penetrate this area. The right panel shows $IV$ characteristics under \SI{6}{GHz} microwave irradiation. The line graphs are at \num{-40}, \num{-30}, \num{-20}, \num{-10} and \SI{0}{dBm} powers for the 250, 500 and \SI{1000}{nm} wide junctions, and are offset in current for clarity. All current plateaus are at multiples of $\Phi_0 f =$\,\SI{12.4}{\micro V}.}
		\label{fig:bi2te3Flux}
\end{figure}

The superconducting leads may be regarded as perfect diamagnets. This leads to flux lines being diverted around the superconducting structure. This causes flux focussing in the junctions, as more flux lines pass through the junctions due to their expulsion from the superconducting bulk. We estimate the amount of flux focussing by considering the shortest distance a flux line has to be diverted to not pass through the superconducting lead. In a long lead, half are passed to the one side and half to the other side. At the end of the lead the flux lines are diverted into the junction area. The flux  diverted is $((W/2-\lambda_L)^2 \times B$, see also the inset of Figure 4. This occurs at both electrodes, and is effectively the same as increasing the junction area by $2\times(W/2-\lambda_L)^2$. Without flux focussing, the expected magnetic field periodicity is given by the dashed line in figure \ref{fig:bi2te3Flux}(a). Correcting for flux focussing and taking $\lambda_L = 39$\,nm results in the solid line, closely describing the measured periods.

The colour graphs in figure~\ref{fig:bi2te3MagneticAndMicrowaveField} show the modulation of the critical current with microwave power. In figure~\ref{fig:bi2te3Flux}(b) $IV$ traces for different applied powers are plotted. The steps all occur at multiples of $\Phi_0 f=$\,\SI{12.4}{\micro V}. A $4 \pi$ periodic Josephson effect will result in steps only at even multiples of $\Phi_0 f$. Shapiro steps are not geometry dependent: in combination with the previously introduced geometry corrected magnetic field periodicity this illustrates the $2 \pi$ periodic Josephson effect in these junctions.

\section{Conclusion}

We investigated superconducting junctions, coupling \nb leads on the surface of a \bite{} flake, by varying the junction width. The critical current and normal state resistance decrease and increase respectively with reduced junction width. However, the \icrn product is found to be geometry dependent, as the normal state resistance does not diverge for zero width. The decreasing \icrn product with reduced junction width is understood when taking into account the resistance due to the entire flake surface. The \icrn product becomes of the order of \num{10} to \SI{15}{\micro V} for wide junctions, similar to previous junctions \cite{Veldhorst2012a}. The junctions are found to be periodic with $\Phi_0$ in a magnetic field when flux focussing is taken into account. Microwave irradiation results in steps at voltages at $k\Phi_0 f$, which is to be expected from junctions with ten to hundred conducting channels contributing to the coupling between the superconducting leads.

Using topological insulators with reduced bulk conductivity should result in increased \icrn products and allow for electrostatic control of the Fermi energy. With similar junction geometries this will  allow for reduction and control of the number of superconducting channels. This step will allow the behaviour of a possible MBS to be uncovered and separated from geometric effects which affect all conducting channels in S-TI-S junctions.

\ack
This work is supported by the Netherlands Organization for Scientific Research (NWO), by the Dutch Foundation for Fundamental Research on Matter (FOM) and by the European Research Council (ERC). 

\section*{References}

\bibliographystyle{iopart-num-nourl}
\bibliography{Bibliography}

\end{document}